\documentclass[aps,prd,twocolumn,nofootinbib,floatfix]{revtex4-1}
\usepackage{setspace}
\usepackage{amsthm}
\theoremstyle{definition}

\usepackage{graphicx}
\usepackage{dcolumn}
\usepackage{multirow}
\usepackage{subfigure}
\usepackage{times,mathptm}
\usepackage{float}
\usepackage{color}
\usepackage{amsmath,amsfonts}
\usepackage{mathptmx}
\usepackage{mathrsfs}
\usepackage{bbm}
\usepackage{bm}
\usepackage{xfrac}

\newcommand{\beq}{\begin{equation}}
\newcommand{\eeq}{\end{equation}} 
\newcommand{\bea}{\begin{eqnarray}}
\newcommand{\eea}{\end{eqnarray}}

\newcommand{\E}{\mathcal{E}}

\renewcommand{\Re}{\text{Re}}

\newcommand{\A}{{\cal A}}

\renewcommand{\d}{\delta}

\renewcommand{\l}{\lambda}

\newcommand{\T}{{\cal T}}
\newcommand{\ta}{\tilde{a}}
\newcommand{\tb}{\tilde{b}}

\renewcommand{\b}{\beta}
\renewcommand{\a}{\alpha}

\renewcommand{\ni}{\noindent}

\newcommand{\vx}{{\vec{x}}}
\newcommand{\vy}{{\vec{y}}}
\newcommand{\vz}{\vec{z}}
\newcommand{\vk}{{\vec{k}}}

\newcommand{\n}{\nu}
\newcommand{\m}{\mu}

\newcommand{\g}{\gamma}
\renewcommand{\r}{\rho}

\newcommand{\s}{\sigma}

\newcommand{\Db}{\mathbf{D}}

\renewcommand{\th}{\theta}

\newcommand{\pbar}{\overline{\psi}}
\newcommand{\vph}{\varphi}
\newcommand{\oh}{\frac{1}{2}}

\newcommand{\dg}{\dagger}
\newcommand{\non}{\nonumber}
\renewcommand{\t}{\tau}
\newcommand{\rf}[1]{(\ref{#1})}
\newcommand{\ra}{\rightarrow}
\newcommand{\pa}{\partial}
\renewcommand{\vec}[1]{\bm #1}

\usepackage{ulem}

\begin{document}

\title{Excited states of massive fermions in a chiral gauge theory} 

\bigskip
\bigskip

\author{Jeff Greensite}
\affiliation{Physics and Astronomy Department \\ San Francisco State
University  \\ San Francisco, CA~94132, USA}
\bigskip
\date{\today}
\vspace{60pt}
\begin{abstract}

\singlespacing
 
     It is shown numerically, in a chiral U(1) gauge Higgs theory in which the left and right-handed fermion components have opposite U(1) charges, that the spectrum of gauge and Higgs fields surrounding a static fermion contains both a ground state and at least one stable excited state.  To bypass the difficulties associated with dynamical fermions in a lattice chiral gauge theory we consider only static fermion sources in a quenched approximation, at fixed lattice spacing and couplings, and with a lattice action along the lines suggested long ago  by Smit and Swift.
     
\end{abstract}

\maketitle

\singlespacing

\section{\label{Intro} Introduction}

    Suppose, in a confining gauge theory with only very massive quarks, we place a static quark and antiquark some large distance apart.  Gauss's law, or equivalently the gauge invariance of physical states, requires that the quark-antiquark pair is associated with a surrounding color electric field, and for a confining gauge theory this is a flux tube state.  But like any quantum system, the flux tube has a ground state and also a spectrum of excited states, and this spectrum has in fact been been observed, in SU(3) pure gauge theory in $D=4$ dimensions, via lattice Monte Carlo simulations \cite{Juge:2002br,Brandt:2016xsp}.  This leads to a natural question:  is there also a non-trivial spectrum for a static fermion-antifermion pair in a non-confining gauge Higgs theory?     By ``non-trivial'' we mean a spectrum containing stable localized excited states, excitations which are distinct from just the ground state plus some number of propagating bosons in the asymptotic particle spectrum of the theory.  
    
    Recent work in both SU(3) gauge Higgs theory \cite{Greensite:2020lmh} and in the abelian Higgs model \cite{Kazue}
 indicates that the answer to this question is affirmative, at least in some range of couplings in the Higgs phase.\footnote{The use of the word ``phase'' in this connection
is deliberate; we contend that there is a meaningful distinction between the confinement and Higgs phases even when the Higgs field is in the fundamental representation.  See ref.\ \cite{Greensite:2020nhg}, which is entirely devoted to this issue.}  However, as far as we know, neither SU(3) gauge Higgs theory nor the relativistic abelian Higgs model describes any system in the real world.   The gauge Higgs theory relevant to particle physics is a chiral gauge theory, with gauge group SU(2)$\times$U(1).  We would like to know whether the quarks and leptons in the electroweak sector of the Standard Model have some spectrum of previously unsuspected \footnote{An exception is the speculation that particle generations might be quantum excitations of a single generation,
cf.\ \cite{Egger:2017tkd}.} excitations, which, being invisible in perturbation theory, would have to be non-perturbative in nature.

    A non-perturbative treatment of a chiral gauge Higgs theory calls for a lattice formulation and there is, up to now, no entirely
satisfactory formulation with a non-abelian gauge group.  Even the abelian version, due to L\"{u}scher \cite{Luscher:1998du}, is very challenging to simulate numerically.  To make progress despite these limitations requires some simplifications. I will work with static fermionic sources in a quenched approximation, with a lattice fermion action proposed long ago by Smit \cite{Smit:1985nu} and Swift \cite{Swift:1984dg}.   I work at fixed lattice spacing and couplings, and do not attempt to take the continuum limit (where the fermions decouple from the gauge field in the Smit-Swift model \cite{Petcher:1993mn}).  Even static fermions propagating only in the time direction have doublers, and we rely on a gauge-invariant Wilson mass term to push up their masses by some finite amount of $O(1)$ in lattice units.

    The aim of this article is to find out whether fermion excitations in gauge Higgs theories, found in refs.\ \cite{Greensite:2020lmh,Kazue}, may also be found  in chiral gauge theories, where the right and left handed fermion components are in different representations of the gauge group.  Of course our ultimate goal is to examine this question in the Standard Model, but we feel that the electroweak sector of Standard Model is too ambitious as a starting point, even with the simplifications already mentioned.  This in part because of the complexity of the fermion sector and the presence of light fermions, and also in part because of the existence of a massless vector boson in the spectrum, which poses some additional technical complications.  To gain some experience with the possible excitations of static fermions in a chiral gauge theory, and to
address the nature of physical states of massive particles in such a theory,  we consider instead the simplest case:  static chiral fermion sources coupled to a U(1) gauge Higgs theory,  where the left and right-handed components have opposite $q=\pm 1$ units of the elementary charge, and the Higgs field carries a single unit of the elementary charge.  
    
\section{The model}

    We begin from the lattice action
\begin{widetext}   
\bea
   S &=& - \b \sum_x \sum_{\m < \n} \Re[U_\m(x) U_\n(x+\hat{\m}) U^*_\m(x+\hat{\n}) U^*_\n(x)]  
       - \g \sum_x \sum_\m \Re[\phi^*(x) U_\m(x) \phi(x+\hat{\m})]  \non \\
       &  & + M \sum_x [\pbar_L(x) \vph(x) \psi_R(x) + \pbar_R(x) \vph^*(x) \psi_L(x)  ] \non \\
      & & - \oh \sum_x \sum_\m  \Bigl[\pbar_R(x), \pbar_L(x)\Bigr] \Db_{\m +}(x)
             \left[ \begin{array}{c}  \psi_L(x+\hat{\m}) \cr \psi_R(x+\hat{\m})\end{array} \right]  
       - \oh \sum_x \sum_\m  \Bigl[\pbar_R(x), \pbar_L(x)\Bigr] \Db_{\m -}(x)
             \left[ \begin{array}{c}  \psi_L(x-\hat{\m}) \cr \psi_R(x-\hat{\m})\end{array} \right] \ ,
\label{S}
\eea
where we impose a unimodular constraint $ |\phi(x)| = 1$ on the Higgs field, defining a double-charged field 
$\vph(x) \equiv \phi^2(x)$, and
\bea
\Db_{\m +}(x) &=& \left[ \begin{array}{cc}
\oh r [\vph^*(x) U_\m(x) + U^*_\m(x) \vph^*(x+\hat{\m})]  &  - \eta^R_\m U_\m^*(x) \cr
- \eta^L_\m U_\m(x) & \oh r[\vph(x) U^*_\m(x) + U_\m(x) \vph(x+\hat{\m})] \end{array} \right] \non \\ \non \\ \non \\
\Db_{\m -}(x) &=& \left[ \begin{array}{cc}
\oh r [\vph^*(x) U^*_\m(x-\hat{\m}) + U_\m(x-\hat{\m}) \vph^*(x-\hat{\m})]  &  \eta^R_\m U_\m(x-\hat{\m}) \cr
\eta^L_\m U^*_\m(x-\hat{\m}) & \oh r[\vph(x) U_\m(x-\hat{\m}) + U^*_\m(x-\hat{\m}) \vph(x-\hat{\m})] \end{array} \right] \ .
\label{D}
\eea
\end{widetext}
Here we have defined
\bea
        \eta^R_k &=& - \eta^L_k = -i\s_k ~~~~~ (k=1,2,3) \non \\
        \eta^R_4 &=&  \eta^L_4 = \mathbbm{1}_2        \ ,
\eea
with  Pauli matrices $\s_k$.

The diagonal entries in the $\Db_{\m \pm}$ matrices are one version of the Wilson mass term, generalized to a chiral
gauge theory.  It is not the only possibility, there are actually infinite possibilites, and one choice which is a little closer to the
original Smit-Swift proposal is to replace the diagonal entries in $\Db_{\m \pm}(x)$ by
\beq
            \left[ \begin{array}{cc}
           r \phi^*(x) \phi^*(x \pm \hat{\m}) & \cr
              &  r \phi(x) \phi(x\pm \hat{\m}) \end{array} \right] \ .
\label{WM2}
\eeq
We will investigate these two possibilities at $r=1$, denoting the Wilson mass term in \rf{D} by WM1  and the second possibility
in \rf{WM2} as WM2.  The $r=0$ case, with no Wilson term, will also be considered for comparison.

  In the continuum a theory of this sort is of course anomalous, and one would have to add additional fermion fields with an opposite assignment of L-R charges to cancel the chiral anomaly.  However, for the present we are interested only in static fermions, there are no fermion loops contributing to an effective action, and the quenched approximation to this model is understood, as already noted.

   In a massive free fermion theory, the mass term
\beq
  m \pbar \psi = m(\pbar_R \psi_L + \pbar_L \psi_R)
\eeq
can be viewed as an interaction vertex which converts a right-handed fermion to a left-handed fermion, and vice versa.  As a result, the physical state corresponding to a massive free fermion is a superposition of the right and left-handed states.  This is the type
of state created in a free field theory by the usual particle/antiparticle creation operators $a^\dg_s(p), \, b_s^\dg(p)$.  With a lattice cutoff and fields with masses much greater than the inverse lattice spacing we consider, motivated by the usual free 
field form, the local operators
\bea
        a_s^\dg(x) &=&  \pbar_{Ls}(x)  + \pbar_{Rs}(x)  \non \\
        a_s(x) &=&  \psi_{Ls}(x)+ \psi_{Rs}(x)  \non \\
        b_s^\dg(x) &=& (-1)^{3-s}(\psi_{L,3-s}(x)  - \psi_{R,3-s}(x) ) \non \\
        b_s(x) &=& (-1)^{3-s}(\pbar_{L,3-s}(x)  - \pbar_{R,3-s}(x) ) \ ,
\label{a}
\eea
where $s=1,2$ is a spin index for the $a,b$ operators, and a Dirac index for the two-component $\psi_L,\psi_R$ fields.
Because static particles propagate only in the time direction we
will only be concerned with the $\m=0$ component of the $\Db_{\m \pm}$ operators, and there are no spin flip terms.  We can therefore arbitrarily choose $a \equiv a_1,  ~b \equiv b_2$, and drop the $s$-indices from here on.

   The operators shown in eqs.\ \rf{a}, in a chiral gauge theory, do not transform covariantly under the U(1) gauge transformations $g(x) = \exp(i\th(x))$, which transform fields according to
\bea
         \psi_L(x) &\ra& g(x) \psi_L(x) ~~~,~~~ \psi_R(x) \ra g^*(x) \psi_R(x) \non \\
         \pbar_L(x) &\ra& g^*(x) \pbar_L(x) ~~~,~~~ \pbar_R(x) \ra g(x) \pbar_R(x) \non \\         
         \phi(x) &\ra& g(x) \phi(x) ~~~,~~~ \vph(x) \ra g^2(x) \vph(x) \non \\
                  U_\m(x) &\ra& g(x) U_\m(x) g^*(x+\hat{\m}) \ .
\eea
If we again view the local ``mass'' term in the action
\beq
S_M = M \sum_x [\pbar_L(x) \vph(x) \psi_R(x) + \pbar_R(x) \vph^*(x) \psi_L(x)  ]  
\label{SM}
\eeq
as a vertex between, e.g., a right-handed fermion and a composite left-handed fermion + Higgs state of the same U(1) charge,  then we may construct $q=\pm 1$ massive fermions from a combination of the corresponding local operators, 
\bea
        \ta^\dg(x) &=&  \pbar_{L}(x) \vph(x)  + \pbar_{R}(x) \non \\
        \ta(x) &=&    \psi_{L}(x) \vph^*(x) + \psi_{R}(x) \non \\
        \tb^\dg(x) &=&  - \psi_{L}(x) \vph^*(x)  + \psi_{R}(x)  \non \\
        \tb(x) &=&  - \pbar_{L}(x)  \vph(x)  + \pbar_{R}(x)  \ .
\label{ab}
\eea 
In the same way one can construct operators transforming covariantly with opposite charge, by combining the right instead of left-handed fermion operators with the squared Higgs field.

\section{\label{pseudomatter} Pseudomatter operators}

   The next step is to create physical states by combining the covariant fermionic operators \rf{ab} with covariant matter (i.e.\ Higgs)
and pseudomatter operators to form gauge invariant operators which act on the ground state of the gauge Higgs theory.
A ``pseudomatter'' operator \cite{Greensite:2017ajx} is an operator which transforms under infinitesimal gauge transformations like a matter field (usually in the fundamental representation of the gauge group), but which is constructed entirely from the gauge field.
We begin with a simple and well known example from continuum electrodynamics in an infinite volume, due originally to Dirac \cite{Dirac:1955uv}. This is the operator
\bea
            \rho(\vx;A) &=& \exp\left[-i {e\over 4\pi} \int d^3z ~ A_i(\vz) {\pa \over \pa z_i}  {1\over |\vx-\vz|}  \right] \ . \non \\
\label{Dirac}
\eea
Let us consider gauge transformations $g(x) = \exp[i\th(x)]$, and we separate out the zero mode $\th(x) = \th_0 + \widetilde{\th}(x)$.
It is easy to verify that under such transformations
\beq
\rho(\vx;g\circ A) = e^{i\widetilde{\th}(x)} \rho(\vx;A) \ .
\eeq
Therefore $\rho$ transforms like a matter field under all infinitesimal gauge transformations, but not under a global U(1) transformation.  We can use this operator to construct physical states containing a single static charge in an infinite volume, e.g.
\beq
           |\Psi_\vx \rangle = a^\dg(\vx) \rho(\vx;A) |\Psi_0 \rangle \ ,
\label{Psix}
\eeq
where $a^\dg$ is the creation operator in \rf{a}.  Due to its invariance under infinitesimal transformations, this state satisfies the Gauss Law constraint, and hence qualifies as a physical state.  It is, in fact, an exact energy eigenstate of free quantum electrodynamics with a static source. 

We note again that while the operator $\ta^\dg(\vx) \rho(\vx;A)$ is invariant under local gauge transformations, it still transforms under global U(1) transformations. This is the hallmark of an operator which can create a physical state associated with a definite isolated charge, given that the system is in a phase such that the ground state is itself an eigenstate of zero charge.   In the case of ordinary electrodynamics, such
charged states do not exist in a finite periodic volume, and this is associated with the fact that the Poisson equation cannot be solved for a point source in a periodic volume.  Intuitively, it is impossible to place a single charge in a periodic volume because there is nowhere for the electric field lines to end.  Generalization of \rf{Psix} to a finite periodic volume requires at least two opposite charges, and this will be relevant to our efforts below.  In a gauge Higgs theory there is also the option of combining $a^\dg, b^\dg$ operators with a Higgs field to create states which are not only gauge invariant, but also charge neutral, and this is the option that was
pointed out long ago, in the context of the Standard Model, by Fr\"{o}hlich, Morcio and Strocchi \cite{Frohlich:1981yi} and
`t Hooft \cite{tHooft:1979yoe} (see also Maas et al.\ \cite{Maas:2017xzh}).   A single charge neutral state can, in fact, exist in a 
periodic volume in any phase, but the use of pseudomatter operators expands the class of quantum states that
can be used to describe physical fermions in a gauge theory, and in fact such operators are required to describe energy eigenstates even in ordinary QED.

   The operator \rf{Dirac}  is only a special case of pseudomatter operators associated with a choice of physical gauge (i.e.\ a gauge free of propagating ghosts).  Let $g_F(x;A)$ be a gauge transformation, in either an abelian or non-abelian gauge theory in either the continuum or the lattice, which takes the gauge field into some physical gauge satisfying a condition $F(A) = 0$, which can be imposed independently on each
time slice.  One may readily verify that under an arbitrary infinitesimal gauge transformation $g(x)$,
\beq
           g_F(\vx,g \circ A) =   g_F(\vx,A) g^\dg(\vx) \ ,
\label{gF}
\eeq
and therefore $g^\dg_F(\vx,A)$ qualifies as a pseudomatter field.  It is also not hard to check that $g_C(\vx;A) = \rho^\dg(\vx,A)$ is
precisely the gauge transformation, in ordinary electrodynamics,  which takes the gauge field $A_\m(x)$ to Coulomb gauge.
The gauge transformation to axial gauge, accomplished by Wilson lines
\beq 
          g_{axial}(x,y,z;A) = P\exp[ie\int_\infty^z dz' A_3(x,y,z')] \ ,
\eeq
is another example, again satisfying \rf{gF}.
Of course not all gauge choices are conditions on the gauge field alone.  In unitary gauge in particular the condition is
$\phi(x)=1$, and
\beq
       g_{unitary}(\vx;\phi) = \phi^*(\vx) \ .
\eeq
In this case the transformation to the gauge is actually a matter, rather than a pseudomatter field.

    In previous work  \cite{Greensite:2020lmh,Kazue}  we have found it useful to employ, for the construction of
physical states containing static charges, a type of pseudomatter operator originally introduced by Vink and Wiese  \cite{Vink:1992ys} in an effort to devise a gauge (the ``Laplacian'' gauge) free of Gribov ambiguities.  On the lattice these operators
are the eigenstates $\zeta_n(\vx;U)$ of the covariant lattice Laplacian operator
\beq
           (-D_i D_i)_{\vx \vy} \zeta_n(\vy;U) = \kappa_n \zeta_n(\vx;U)  \ ,
\eeq
where
\bea
 (-D_i D_i)_{\vx \vy} = 
    \sum_{k=1}^3 \left[2  \d_{\vx \vy} - U_k(\vx) \d_{\vy,\vx+\hat{k}}  - U_k^{\dg}(\vx-\hat{k}) \d_{\vy,\vx-\hat{k}}   \right] \ , \non \\  
\label{Laplacian}
\eea
defined at fixed time on a $D=4$ dimensional periodic lattice.\footnote{These eigenstates are computed numerically via the Arnoldi algorithm, as implemented in the ARPACK software package (https://www.caam.rice.edu/software/ARPACK/).}   Like $\r(\vx;A)$, the 
$\zeta_n(\vx;U)$ transform covariantly like charge $q=1$ matter fields, under infinitesimal gauge transformations, but are
invariant under global U(1) transformations.

     We then consider constructing a set of $N=2n_{ev}+1$ physical fermion-antifermion states using the first $n_{ev}$ Laplacian eigenstates with lowest $\kappa_n$ as follows (no sum over $i$):
\beq
        \Phi_i(R;\psi,\pbar,U) = \{\ta^\dg(\vx) \xi^*_i(\vx;U)\} ~ \{\tb^\dg(\vy) \xi_i(\vy;U) \} ~\Psi_0(U)  \ ,
\label{Ups}
\eeq
where $\Psi_0(U)$ is the ground state of the pure gauge theory, and where
\beq
             \xi_i(\vx;U) = \left\{ \begin{array}{cl} 
                                   \zeta_i(\vx;U) & i \le n_{ev} \\
                                   \vph(x) \zeta^*_{i-n_{ev}}(\vx;U) & n_{ev}+1 \le i \le 2n_{ev} \\
                                   \phi(\vx) & i=2n_{ev} + 1 \end{array} \right. \ .
\eeq
 The pseudomatter fields with index $n_{ev}<i \le 2n_{ev}$ allow for the possibility mentioned below \rf{ab}, that the 
$\ta, \, \tb$ operators could have be chosen to transform with the opposite charge.  All fermionic expectation values are
computed in the local measure
\beq
        D\pbar_R D\pbar_L D\psi_R D\psi_L ~ e^{-S_M} \ ,
\label{measure1}
\eeq
\ni where $S_M$ was defined in \rf{SM}.  In this measure
\bea
         \langle \psi_L(x) \pbar_R(y) \rangle &=& {1\over M} \d_{xy} \vph(x) \non \\
         \langle \psi_R(x) \pbar_L(y) \rangle &=& {1\over M} \d_{xy} \vph^*(x) \ .
\eea

     The bra state corresponding to \rf{Ups}
\beq
      \langle \Phi_i | =  \langle \Psi_0(U)| ~  \{\tb(\vy) \xi^*_i(\vy;U) \} \{\ta(\vx) \xi_i(\vx;U)\}   
\eeq
is obtained by the replacements
\bea
 \ta^\dg(\vx) \xi^*_i(\vx;U) &\ra& \ta(\vx) \xi_i(\vx;U) \non \\
 \tb^\dg(\vy) \xi_i(\vy;U) &\ra&  \tb(\vy) \xi^*_i(\vy;U) \ .
 \eea
 While these replacements might seem obvious from the point of view of
 canonical quantization, they are a little less obvious in the Euclidean formulation, where $\pbar_{L,R}$ and $\psi_{L,R}$ are independent variabes.  Nevertheless, one can check that this is what is required in order that states have positive norms in the measure \rf{measure1}.  Likewise, it is necessary that $\ta, \tb$ in the ket go over to $-\ta^\dg, -\tb^\dg$ in the bra.   In more
 generality, the correspondence required by positivity in the measure \rf{measure1}, between operators in the ket and operators
 in the bra is
 \bea
 \begin{array}{ccc}
    \underline{\text{ket}}   &   &   \underline{\text{bra}} \cr
    \pbar_R    &\ra&  \vph^* \psi_L \cr
    \pbar_L    &\ra&  \vph \psi_R \cr
    \psi_R    &\ra&  -\pbar_L \vph \cr
    \psi_L    &\ra&  -\pbar_R \vph^*  \ .
\end{array} 
\eea

    One might ask what is wrong, in a non-confining theory, with a state containing a single charge, e.g.
\beq
             \Phi^{single}_i(R)  = \ta^\dg(\vx) \zeta^*_i(\vx;U) \Psi_0 \ .
\label{single}
\eeq
The answer is that a state of this sort doesn't propagate, i.e.
\beq
 \langle \Phi^{single}_i(R)| \T  |\Phi^{single}_i(R) \rangle = 0 \ .
\label{singleT}
 \eeq
 The reason for this is that the Laplacian eigenstates $\zeta_i(\vx,U)$ contain, apart from their space variation, a global, $\vx$-independent but field-dependent  phase factor $\exp(i \gamma_i[U])$ which depends on $U$ on a time-slice.  The left hand side of
 \rf{singleT} vanishes due to wild oscillations of this phase factor.  The existence of this global phase factor was recognized already in the original work of Vink and Wiese \cite{Vink:1992ys}.  For the same reason, a state of the form
 \beq
             \Phi_{ij}(R)  = \{\ta^\dg(\vx) \zeta^*_i(\vx;U)\} ~  \{\tb^\dg(\vy) \zeta_j(\vy;U) \} ~ \Psi_0 (U)
\eeq
with $i \ne j$ will also not propagate. Only for $i=j$ do the global phase factors cancel out on a time slice.  So we will restrict the calculation to a set of states of the form \rf{Ups}.

\section{\label{search} Searching for excited states}

   The problem of finding the first excited state of the fermion-antifermion system is simplified by the fact that, in this particular
model, at couplings $\b=3.0, \g=1.0$, one of the states in the subspace spanned by the $\{ \Phi_i(R) \}$ at each $R$ turns out
to be very nearly an exact energy eigenstate (or, more precisely, an eigenstate of the transfer matrix).  This means that
states which are constructed to be orthogonal to this state will all be dominated, after a sufficient interval of evolution in Euclidean time, by the first excited state.  

   Let $\t = e^{-H}$ be the transfer matrix, where $H$, which is formally just the logarithm of $\t$, would be the Hamiltonian operator
in the continuous time limit, and let $\T = e^{-(H-\E_0)}$ be the transfer matrix divided by its highest eigenvalue $e^{-\E_0}$, where
$\E_0$ is the ground state energy of the vacuum in the absence of static sources.  From here on, in a slight departure from normal terminology, I will refer to the rescaled operator $\T$ as simply the ``transfer matrix,'' since it is the matrix elements of this rescaled operator which we actually compute numerically.

     We restrict attention to the Hilbert space containing a static fermion-antifermion pair a distance $R$ apart.  The non-orthogonal states $\{ |\Phi_i \rangle \}$, defined in the last section, span a tiny $N=2n_{ev}+1$ dimensional subspace of this Hilbert space, with a matrix of overlaps denoted
\beq
           [O]_{ij}(R) = \langle \Phi_i(R)| \Phi_j(R) \rangle \ .
\label{O}
\eeq
We also define
\beq
           [\T^T]_{ij}(R) = \langle \Phi_i(R)|\T^T | \Phi_j(R) \rangle \ ,
\label{T}
\eeq
where $T$ refers to a number of time steps in Euclidean time.  Both $[O]$ and $[\T^T]$ are finite $N\times N$ matrices, and we are interested in diagonalizing the operator $\T^T$ (transfer matrix to the power $T$) in the small subspace of Hilbert space spanned by the non-orthogonal states $\{ |\Phi_i \rangle \}$.  This is achieved, at each $R,T$, by solving the generalized eigenvalue problem
\beq
         [\T^T](R) \vec{\upsilon}^{(n)}(R,T) = \lambda_{n}(R,T) [O](R) \vec{\upsilon}^{(n)}(R,T) \ .
\label{GEP}
\eeq
We then have a set of $N=2n_{ev}+1$ orthogonal states
\beq
 \Psi_n(R,T) = \sum_{i=1}^N \upsilon^{(n)}_{i}(R,T) \Phi_i(R,T) \ ,
\eeq      
with the property that
\beq
          \langle \Psi_m(R,T) | \T^T | \Psi_n(R,T) \rangle = \l_n(R,T) \d_{nm} \ ,
\eeq
despite the fact that in general the $\Psi_n(R,T)$ are not eigenstates of the transfer matrix in the full Hilbert space.
Let $\{ \Omega_\a(R) \}$ denote the exact eigenstates of the transfer matrix, with eigenvalues $\exp[-E_\a]$ in the Hilbert space
containing the two static charges.  Then we have
\beq
            \T^T = \sum_\a  | \Omega_\a \rangle \langle \Omega_\a | e^{-E_\a T}
\eeq
and
\bea
     \l_n(R,T) &\equiv& \langle \Psi_n(R,T)| \T^T| \Psi_n(R,T) \rangle \non \\
     &=& \sum_\a |\langle \Psi_n| \Omega_\a \rangle|^2 e^{-E_\a T} \non \\
     &=& \sum_\a a_{n\a}(R,T) e^{-E_\a T} \ .
\label{Tnn}
\eea 
Now at large $T$, taking $\Omega_1(R)$ to be the ground state of the system containing the two static fermions, this matrix element would seem to converge to
\beq
         \l_n(R,T) \ra  a_{n1}(R,T) e^{-E_1(R) T} \ ,
\eeq
and in this way (assuming that $a_{n1}(R,T)$ converges to a constant), we could determine only the ground state energy $E_1(R)$
from a logarithmic plot of  $\l_n(R,T)$ at large $T$.  But this assumes that all the $a_{n1}$ are non-zero.  Suppose instead that
 \beq
 \Psi_1(R,T) \approx \Omega_1(R,T) ~~~,~~~ a_{n1} \approx \d_{n1} \ .
 \label{approx}
 \eeq
 Then
\beq
\l_n(R,T) \ra a_{n2}(R,T) e^{-E_2(R) T} ~~~ \text{for} ~~n \ne 1 \ .
\eeq
If the approximate equality in \rf{approx} is close enough to an exact equality, then we may in principle extract the ground state energy $E_1$ from $\l_1(R,T)$, and the energy of the first excited state from the large Euclidean time behavior of {\it any} of 
the $\l_{n>1}(R,T)$.  

 It must be noted in passing that the connection between energy and Euclidean time evolution depends on the positivity of the transfer matrix.  To the author's knowledge this positivity has never been proven for the Smit-Swift model, or for other attempts to put chiral gauge theories on the lattice.  That is an important question of principle, but we will ignore it for now.  It does not come up in the
results to be presented next.    \\

\section{The Calculation}

    In order to carry out the search for excited states as described in Section \ref{search}, we need to compute numerically the matrices shown in \rf{O}, \rf{T}, and then solve the generalized eigenvalue problem \rf{GEP}.   Expressed as Euclidean
vacuum expectation values, these matrices are
\bea
[\T^T]_{ji} &=& \Bigl\langle \tb(\vy,t+T) \xi^*_j(\vy,t+T;U) \ta(\vx,t+T) \xi_j(\vx,t+T;U) \non \\
& &  \times \ta^\dg(\vx,t) \xi^*_i(\vx,t;U)  \tb^\dg(\vy,t) \xi_i(\vy,t;U) \Bigr\rangle \ ,
\eea
and
\bea
[O]_{ji} &=& \non \Bigl\langle \tb(\vy,t) \xi^*_j(\vy,t;U) \ta(\vx,t) \xi_j(\vx,t;U) \non \\
& &  \times \ta^\dg(\vx,t) \xi^*_i(\vx,t;U)  \tb^\dg(\vy,t) \xi_i(\vy,t;U) \Bigr\rangle  \non \\  \ .
\eea

We expand in a Taylor series the terms in $e^{-S}$ involving $D_{\m \pm}$ and integrate over fermion fields in the measure 
\rf{measure1}, keeping only terms at leading order in the hopping expansion and discarding, since we work only in the quenched approximation, all fermion loops.  We define
\bea
          q^{0-}_{ji}(\vx,t) &=& 2\xi_i(\vx,t;U) \xi_j^*(\vx,t;U)  \non \\
          q^{0+}_{ji}(\vx,t)  &=& 2\xi_i^*(\vx,t;U) \xi_j(\vx,t;U)  \ ,
\eea
and 
\begin{widetext}                       
\bea
          Q^{+T}_{ji}(\vx,t) &=& \Bigl[\xi_i(\vx,t),-\vph^*(\vx,t)\xi_i(\vx,t) \Bigr] ~\Db_{4+}(\vx,t) 
         \left( \prod_{\tau=1}^{T-1} {\mathbf F}(\vx,t+\tau)\Db_{4+}(\vx,t+\tau) \right) \left[ \begin{array}{c} 
                           \vph(\vx,t+T) \xi_j^*(\vx,t+T)  \\
                            -\xi_j^*(\vx,t+T)  \end{array} \right]  \non \\ \non \\
          Q^{-T}_{ji}(\vx,t) &=& \Bigl[\xi_j(\vx,t+T),\vph^*(\vx,t+T)\xi_j(\vx,t+T) \Bigr][ ~ \Db_{4-}(\vx,t+T) 
         \left( \prod_{\tau=1}^{T-1} {\mathbf F}(\vx,t+T-\tau) \Db_{4-}(\vx,t+T-\tau) \right) \left[ \begin{array}{c} 
                           \vph(\vx,t) \xi_i^*(\vx,t)  \\
                            \xi_i^*(\vx,t)  \end{array} \right]  \non \\ \ ,
\label{QT}
\eea 
\end{widetext}
where
\beq
           {\mathbf F}(x) = \mbox{diag}[\vph(x),\vph^*(x)] \ .
\eeq
    Then the overlap matrix is 
\bea
           [O]_{ji}(R) &=& \langle \Phi_j |\Phi_i \rangle \non \\
                       &=& \langle q^{0-}_{ji}(\vx,t) q^{0+}_{ji}(\vy,t) \rangle \ ,
\label{olapmat}
\eea    
and to leading order in $1/M$ (i.e.\ the leading term in the hopping parameter expansion)
\bea
          [\T^T]_{ji}(R) &=& \langle \Phi_j |  \T^T | \Phi_i \rangle \\
                               &=& \langle  Q^{+T}_{ji}(\vx,t)  Q^{-T}_{ji}(\vy,t)   \ .
\label{Tmat}
\eea
These are vacuum expectation values of operators which depend only on the gauge and scalar field variables, obtained from   numerical simulation of U(1) gauge Higgs theory.  From those expectation values we obtain $\l_{n}(R,T)$ by the
method described in Section \ref{search}.

    It should be noted that in (\ref{olapmat},~\ref{Tmat}) we have dropped powers of the hopping parameter $1/M^2$.  When included, these simply add a constant of $2M$ to the energy of the static fermion-antifermion pair, in each energy eigenstate.

\section{Results}

\begin{figure}[t!]
 \includegraphics[scale=0.6]{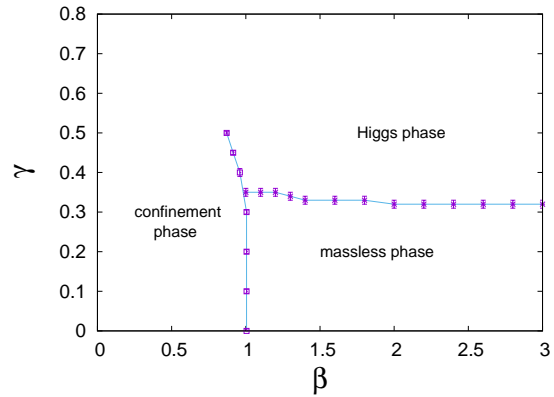}
 \caption{Phase diagram of the $q=1$ abelian Higgs model.}
  \label{thermo}
 \end{figure}

    The first step, before computing excitation energies, is to map out the phase diagram of the action in eq.\ \rf{S}, dropping the
fermionic terms.   That action is simply the abelian Higgs model with a unimodular Higgs field of unit charge, and the phase diagram
of that theory was approximately determined long ago via lattice Monte Carlo by Ranft et al.\ \cite{Ranft:1982hf}, on lattices that were tiny and by numerical methods that were a little primitive by today's standards.  A more accurate determination is shown
in Fig.\ \ref{thermo}.  The transition line between the massless and Higgs phases is associated with a ``kink'' in a plot of the link
action $\langle \mbox{Re}[\phi^*(x) U_i(x) \phi(x+ \hat{i})]\rangle$ vs.\ $\g$ at fixed $\b$, as observed previously in the abelian Higgs
model with a double-charged $q=2$ Higgs field \cite{Matsuyama:2019lei}.  The remaining transition lines were determined from the peaks of lattice volume dependent link and plaquette susceptibilities.\footnote{I am grateful to Kazue Matsuyama for supplying me with the data for this figure.}  I would mention here that although the confinement and Higgs
regions are not entirely separated by a thermodynamic transition, they are in fact different phases, characterized by different types of confinement (``separation-of-charge'' and color confinement, respectively),  and distinguished by the spontaneous breaking of global center symmetry, as detected by an order parameter analogous to the Edwards-Anderson order parameter for a spin glass.  For a detailed discussion of this point, cf.\  \cite{Greensite:2020nhg}, where the transition line is computed for the SU(2) gauge Higgs model.  For the $q=1$ abelian Higgs model with a Higgs field of unit charge, the transition line separating the confinement and spin glass (i.e.\ Higgs) phases has not yet been determined.

     We look for excitations in the spectrum of static fermions at two points in phase diagram.  The first, at $\b=3.0, {\g=1.0}$ is
deep in the Higgs phase.  The second, at $\b=3.0, {\g=0.5}$, while still in the Higgs phase, is somewhat closer to the massless to Higgs phase transition at $\b=3.0, \g=0.32$.  We will find some qualitative differences in these two cases but do not attempt, in this article, a systematic study of the situation throughout the Higgs phase. 

    In this work the number of Laplacian eigenstates is $n_{ev}=4$, so there are nine states $\Phi_{1-9}(R)$ at each fermion-antifermion separation $R$.

   The mass gap of the theory in the Higgs phase is extracted from correlators of gauge invariant operators
\beq
           \A_i(x) = \mbox{Im}[\phi^*(x) U_i(x) \phi(x+ \hat{i})]  \ ,
 \eeq
 which can be recognized as the lattice 4-vector potential $A_i(x)$  in unitary gauge.  We construct the zero-momentum
 operator
 \bea
           \A_i(\vk=0,t) = {1\over L^3} \sum_\vx \A_i(\vx,t)  \ ,
 \eea
 
 \ni and the correlator \footnote{Suitably modified for time periodicity.  Of course, since any axis can be defined as the ``time'' axis, we make can make use of both time translation symmetry and hypercubic symmetry on the lattice to increase our statistics.}
 \bea
            G(T) =  {1\over 3}  \sum_{i=1}^3 \langle \A_i(\vk=0,t) \A_i(\vk=0,t+T) \rangle \non \\  \ .
\label{G}
\eea
 
\begin{figure}[htpb]
 \includegraphics[scale=0.6]{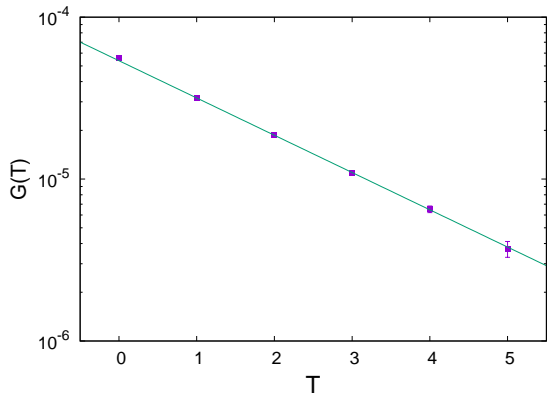}
 \caption{The $G(T)$ gauge field correlator vs.\ $T$, where $G(T)$ is defined in \rf{G}.  The correlator was computed for 
 ${\b=3, \g=1.0}$ on a $16^4$ lattice volume. On this log plot, the photon mass $m=0.530(1)$ 
 is obtained from the slope of a straight-line fit to the data.}
 \label{photon}
 \end{figure}

The results for $G(T)$ vs.\ $T$, obtained at $\b=3.0$ and ${\g=1.0}$ on a $16^4$ lattice volume, are shown on a logarithmic plot in Fig.\ \ref{photon}.  Fitting the data to a simple exponential falloff, the photon mass in lattice units is
determined to be $m_{photon} = 0.530(2)$ at $\g=1$.

 \subsubsection{Remarks}
 
     It may be helpful to make some general remarks before proceeding to the numerical results.  What will be seen in those
results is that $\Psi_1$ is very close to an exact eigenstate of the transfer matrix, and, by construction, the $\Psi_{n>1}$ are orthogonal, or very nearly orthogonal, to $\Psi_1$.  The qualification ``nearly'' comes from slight numerical inaccuracies in the solution of the generalized eigenvalue problem, which is carried out by standard Matlab software.  The $\Psi_{n>1}$ are not eigenstates of the transfer matrix, at least not outside the truncated space spanned by the $\Phi_i$.  However, because of their (near) orthogonality to $\Psi_1$, which {\it is} an eigenstate in the full Hilbert space, the evolution of the $\l_{n>1}(R,T)$ will be
dominated by energy eigenvalues of the full theory which differ from $E_1$, as explained in section \ref{search}.  At $\g=1$, $\Psi_1$ is the ground state, and is essentially, for $n_{ev}=4$, the (appropriately normalized) state $\Phi_9$.  The situation is a little different, as we will see, at $\g=0.5$, which is closer to the massless phase.

    The numerical results reported below were obtained on a $14^3 \times 22$ lattice volume, with data taken on 600 lattices.
These were obtained from 10 independent runs, with data taken (after 2000 thermalization sweeps) on 60 lattices separated
by 200 update sweeps. Error bars are obtained as the standard error of the mean derived from the ten separate runs. In all cases we employ $n_{ev} = 4$ Laplacian eigenstates.  In the Wilson mass terms WM1 (in \rf{D}) and WM2 defined in \rf{WM2} we set $r=1$.

\subsection{Higgs phase at $\b=3, \g=1.0$}  

 \subsubsection{Wilson term WM1}

\begin{figure}[htpb]
\subfigure[~]{
 \includegraphics[scale=0.5]{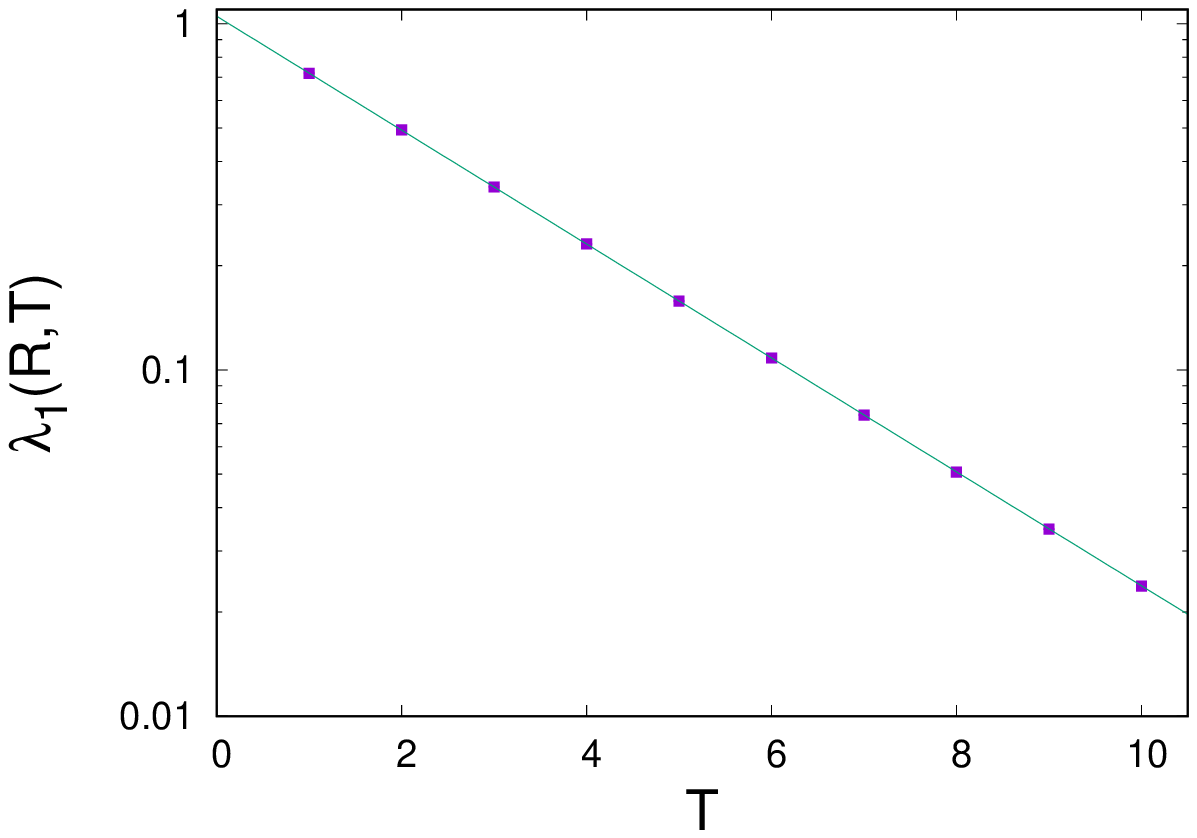}
\label{zx1g10}
}
\subfigure[~]{
 \includegraphics[scale=0.5]{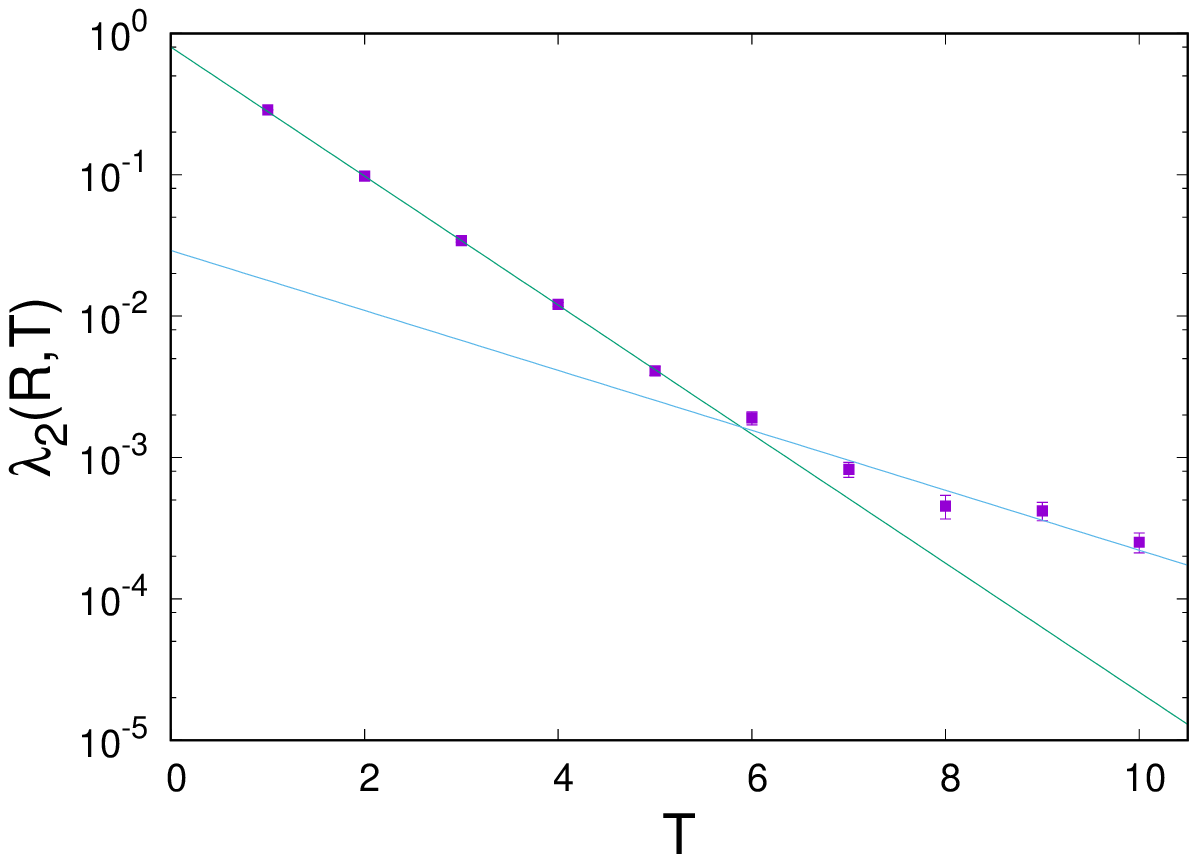}
\label{zx2g10}
}
\subfigure[~]{
\includegraphics[scale=0.5]{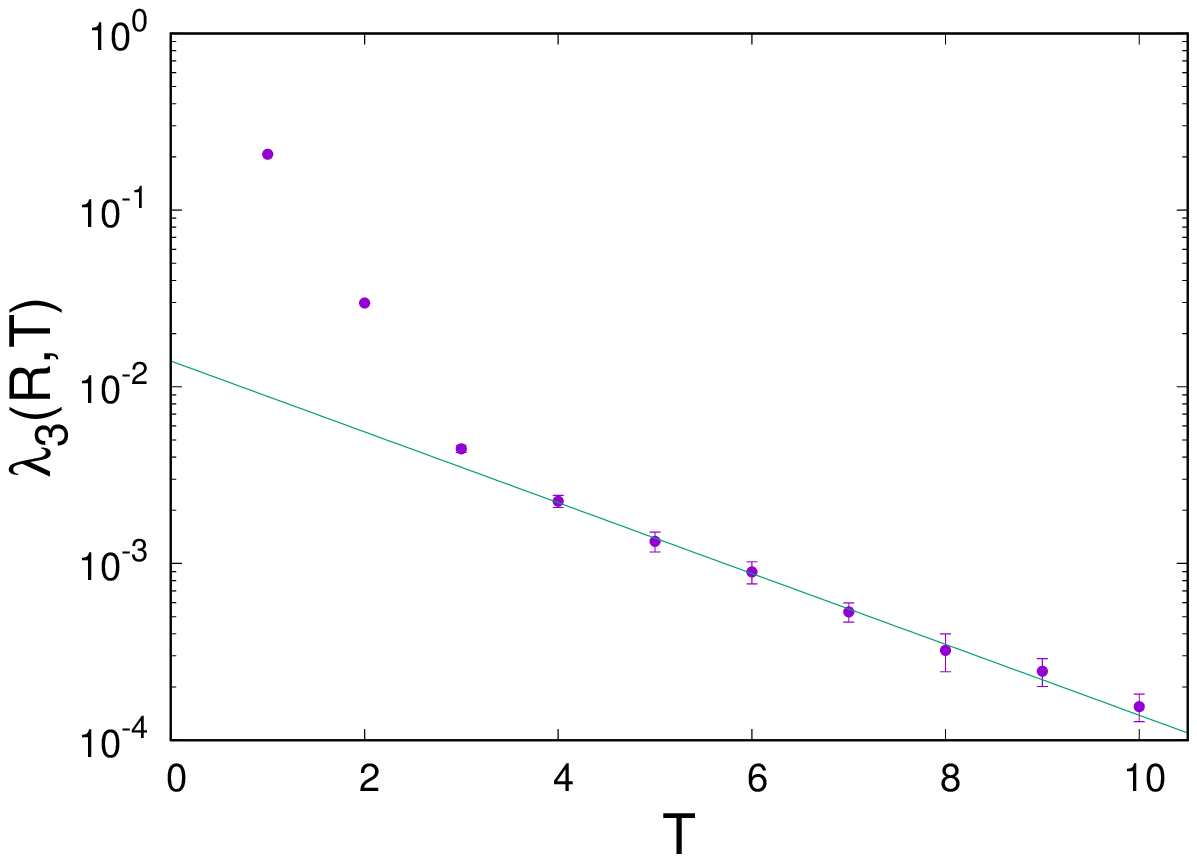}
\label{zx3g10}
} 
\subfigure[~]{
 \includegraphics[scale=0.5]{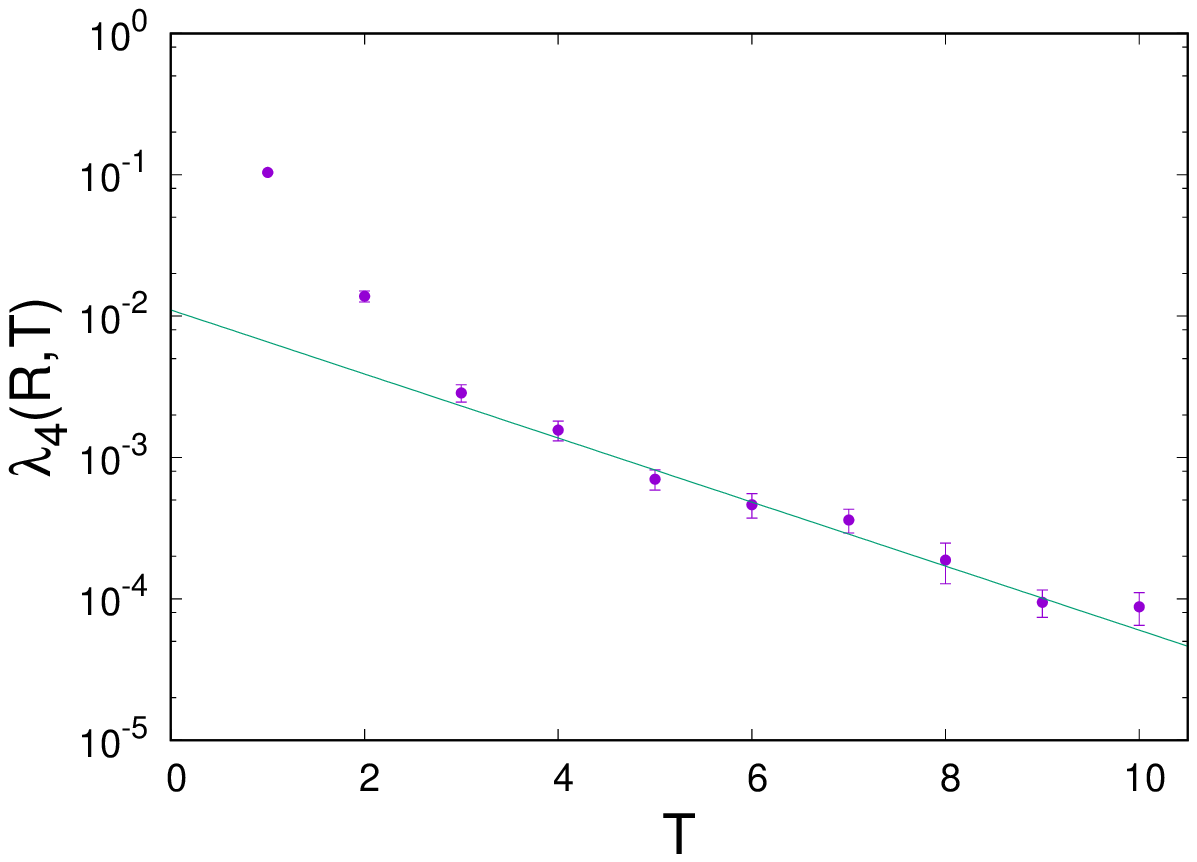}
\label{zx4g10}
}
\caption{$\l_n(R,T)$ vs.\ $T$ for the WM1 non-local mass term, at $R=4.243$ and $\g=1.0$. (a) The ground
state energy $E_1$ is derived from a fit to $\l_1(R,T)$ over the full range of $T$ (error bars are smaller than the symbol size).  (b) The energy $E_3$ is extracted from 
a fit to $\l_2(R,T)$ in the range $T=1-5$. (c) The energy $E_2$ is derived from a fit to $\l_3(R,T)$ in the range $T=4-10$.
(d) Energies $E_2$ can also be extracted from a fit to $\l_4(R,T)$ at large $T$, and these are consistent with those derived from
$\l_3$, albeit with larger $\chi^2$ and error bars.}
\label{zxg10}
\end{figure}

   We begin with the fact that $\Psi_1$ is a near-exact eigenstate of the full transfer matrix.  The justification is seen in a log plot, 
Fig.\  \ref{zx1g10}, of $\l_1(R,T)$ vs.\ $T$, where $R$, in this example, is ${R=4.243}$.  There is nothing special about this particular choice of $R$, and we have very similar plots at all $R$ values, but for the sake of comparison we will stick to this choice in all subsequent plots of $\l_n(R,T)$ shown in this and the following subsections. The data fits a straight line on a log plot over the full range of Euclidean time, starting at $T=1$,
and extrapolates to $1.05$ as $T\ra 0$.  This can only be true if $\Psi_1(R,T) \approx \Omega_1(R)$ for all $T$. The next obvious question is whether this ground state is simply the neutral state
obtained by multiplying fermion operators by the scalar field, a state which is independent of the pseudomatter fields, and for
$n_{ev}=4$ this is the state $\Phi_9$.
The answer to the question is yes.  One can easily compute the overlap, at any $R$, of the ground state with the neutral state
\bea 
          f &=& {|\langle \Phi_9(R)|\Psi_1(R,T) \rangle|^2 \over \langle \Phi_9(R)|\Phi_9(R) \rangle} \non \\
            &=& {1\over 4} \left|\sum_{i=1}^9 \upsilon_i^{(1)} [O]_{9i} \right|^2 \ ,
\label{f}
\eea
where $[O]_{ij}$ and $\upsilon_i^{(n)}$ were defined in eqs.\ \rf{O} and \rf{GEP}.  The result, at all $R,T$, is that the overlap is within
0.1\% of unity.  This far into the Higgs phase, the ground state is created by precisely the type of operator suggested long ago by Fr\"{o}hlich, Morcio and Strocchi \cite{Frohlich:1981yi} and by `t Hooft \cite{tHooft:1979yoe}.  What is new, as we will now see, is that this is not the only stable and localized quantum state of the system.  In some regions of the phase diagram it may not even be
the ground state.

   In Fig.\ \ref{zx2g10} we display a plot of  $\l_2(R,T)$, again at ${R=4.243}$.  The first five data points are fit fairly well by a straight line on a log plot, which extrapolates to only a little below 1.0 at $T=0$.  So the data in this range seems to be dominated by a single
eigenstate of the transfer matrix, with a slope corresponding to an energy $E_3 = 1.064(4)$.  But $\Psi_2$ is not an exact eigenstate,
and at higher $T$ a small admixture of the first excited state becomes dominant.  The corresponding energy is not so easy to
extract from only the data in the range $T  \ge 6$, and the precise value is rather sensitive to the range of the fit.  However, if
we do make a fit to the range $T=6-10$,  the energy comes out to be, rather imprecisely, $E_2 = 0.49(8)$.

   Fortunately, the first excitation energy $E_2$ can be determined more accurately from $\l_3(R,T)$, as seen in Fig.\ \ref{zx3g10}, same $R$ as before.  A fit to the data  in the range $T=4-10$ yields $E_2=0.45(1)$.  Of course, as stated previously, the energy
of the first excited state could be determined in principle from the large-$T$ behavior of any $\l_{n>2}$.  Figure \ref{zx4g10} 
shows a fit to $\l_4(R,T)$ in the same $T=4-10$ range.  The fit is not as good, the error bar is larger, but the
energy comes out to be ${E_2 = 0.50(6)}$, compatible with the value derived from $\l_3$.  Generally the $\l_3(R,T)$ values give
a more accurate estimate for $E_2$ as compared to $\l_4(R,T)$, but there are a few isolated exceptions at some separations $R$, in which the $\chi^2$ value of the fit to $\l_3$ is much larger than the corresponding fit for $\l_4$, and in those few cases
we use the latter fit.

\begin{figure}[htbp]
 \includegraphics[scale=0.6]{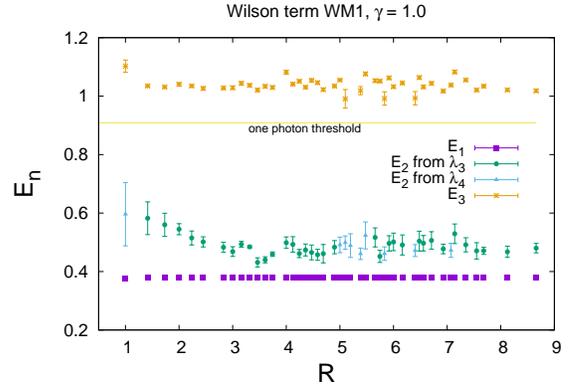}
 \caption{Energies $E_1,E_2,E_3$ vs.\ $R$ at $\b=3, \g=1$, shown together with the one photon threshold.  The Wilson mass term is WM1.}
  \label{ERzxg10}
 \end{figure}

   The values for $E_{1,2,3}(R)$ obtained by these methods are shown in Fig.\ \ref{ERzxg10}.  The line denoted ``one photon threshold'' is obtained by adding the photon mass of $m_{ph}=0.530$ in lattice units to the ground state energy $E_1=0.379$.  We see that $E_3$ is above this threshold, so it is probably best understood as the ground state plus one photon.  But the energy $E_2$
is well below that threshold.  It means that a ground state + one photon interpretation for this excited state is untenable.  
$E_2$ is the energy of a {\it stable} excited state.  It cannot decay to the ground state by photon emission.

 \subsubsection{Wilson term WM2}

\begin{figure}[htbp]
\includegraphics[scale=0.6]{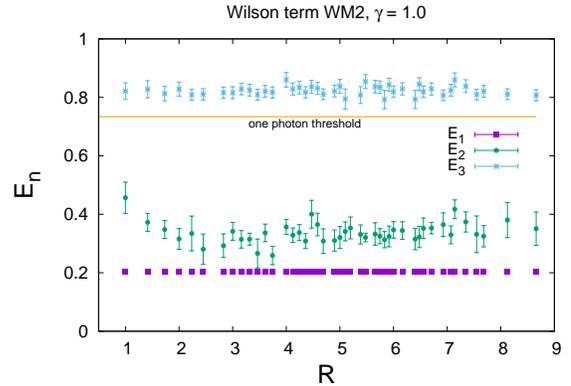} 
\caption{Same as Fig.\ \ref{ERzxg10}, but for the WM2 Wilson term.}
\label{ERsxg10}
\end{figure}

The analysis of data obtained with an action containing the Wilson term WM2 \rf{WM2} is identical to the analysis described in the previous subsection.  The energies are slightly different numerically, as would be expected away from the continuum limit, but everything else is qualitatively the same.  The energy eigenvalues are displayed in Fig.\ \ref{ERsxg10}.  The first excited state, lying
well below the one photon threshold, is stable.

\subsubsection{no Wilson term, $r=0$}
 
\begin{figure}[t!]

\subfigure[~]{
 \includegraphics[scale=0.5]{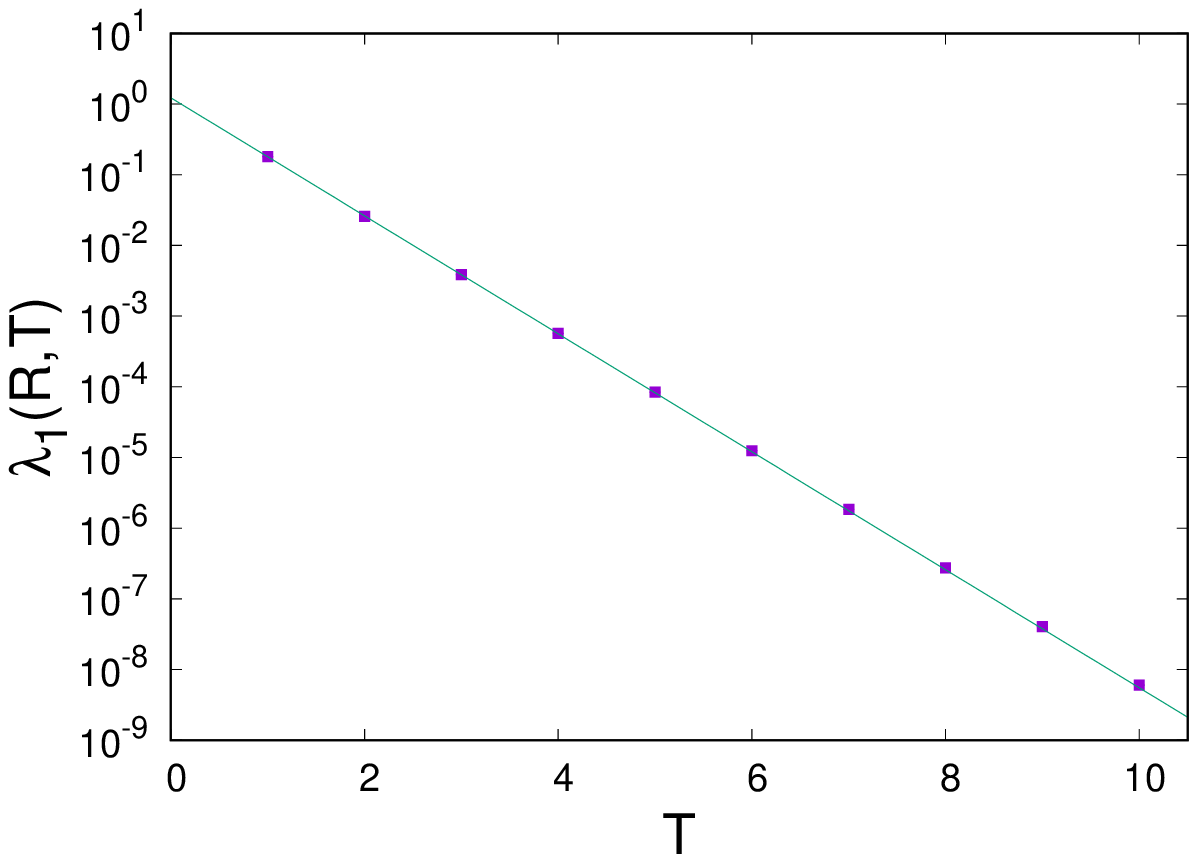}
\label{rx1g10}
}
\subfigure[~]{
 \includegraphics[scale=0.5]{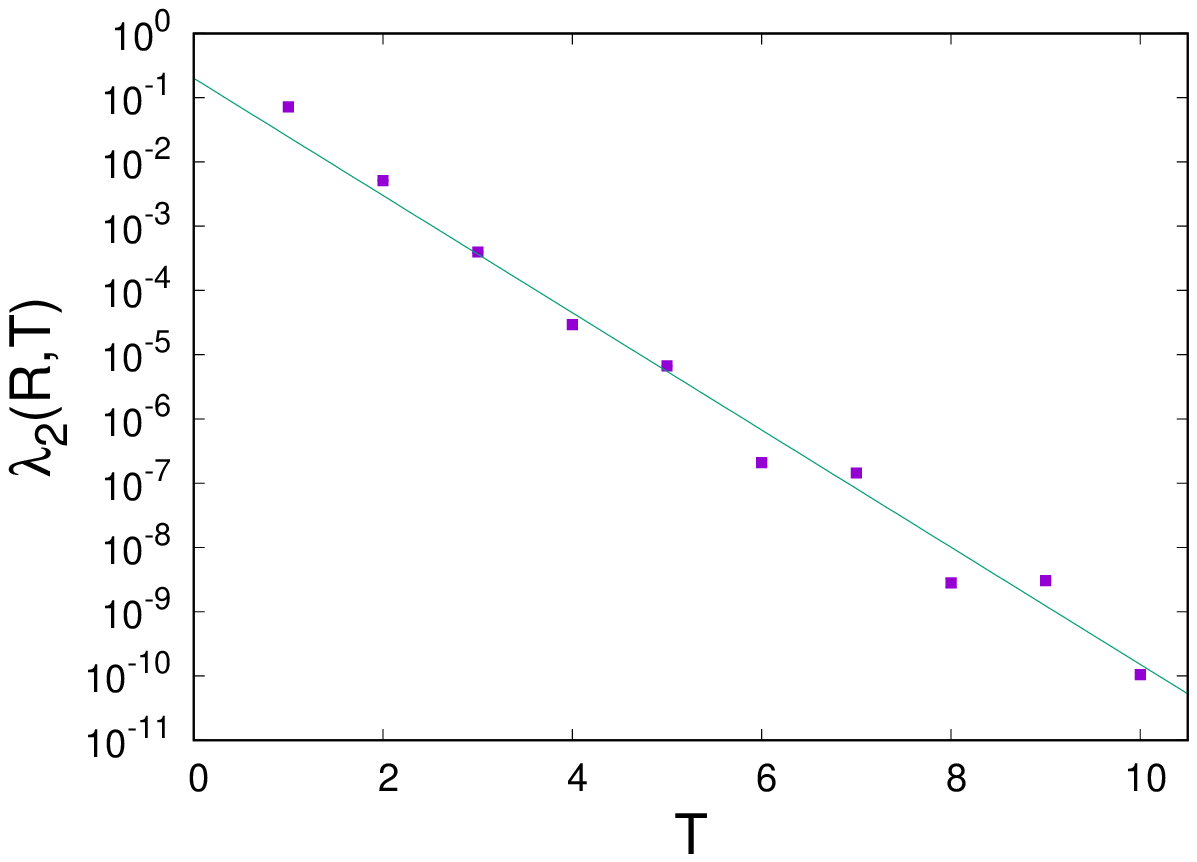}
\label{rx4g10}
}
\subfigure[~]{
 \includegraphics[scale=0.5]{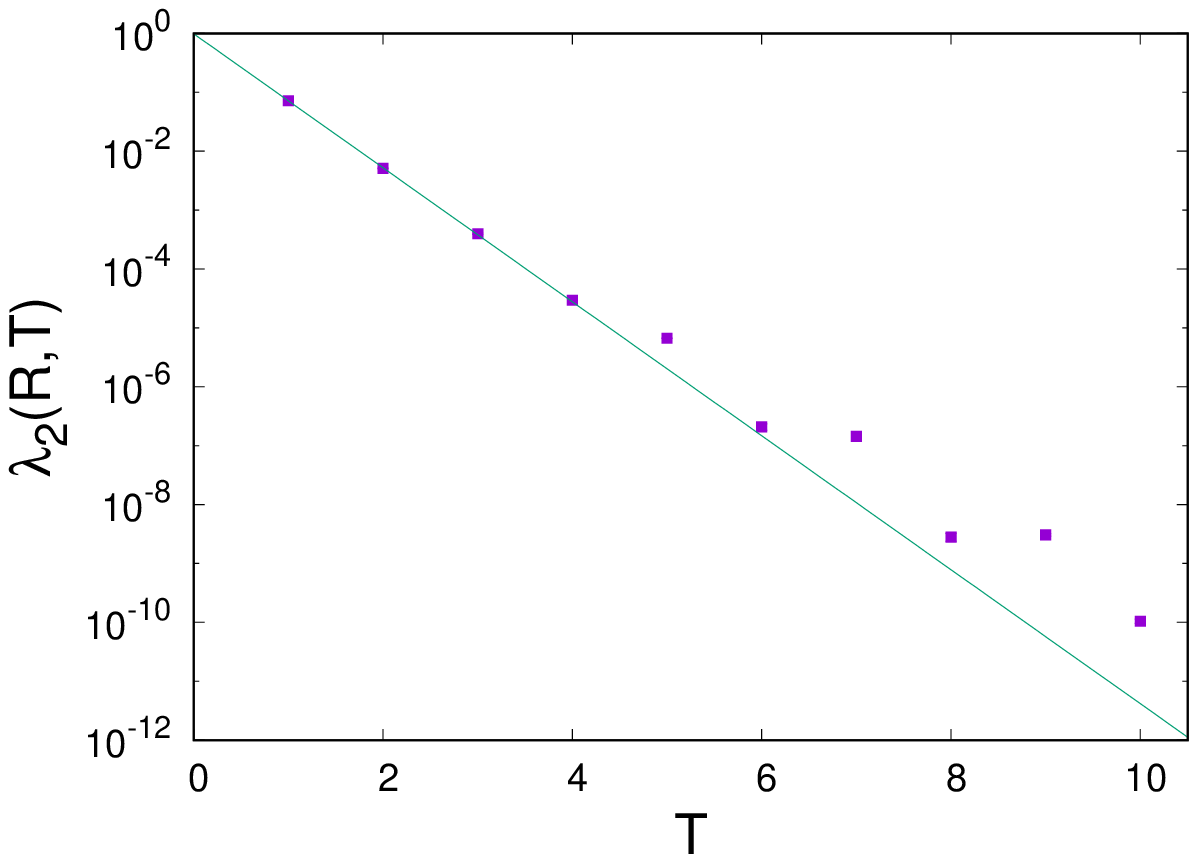}
\label{rx3g10}
}
\subfigure[~]{
 \includegraphics[scale=0.5]{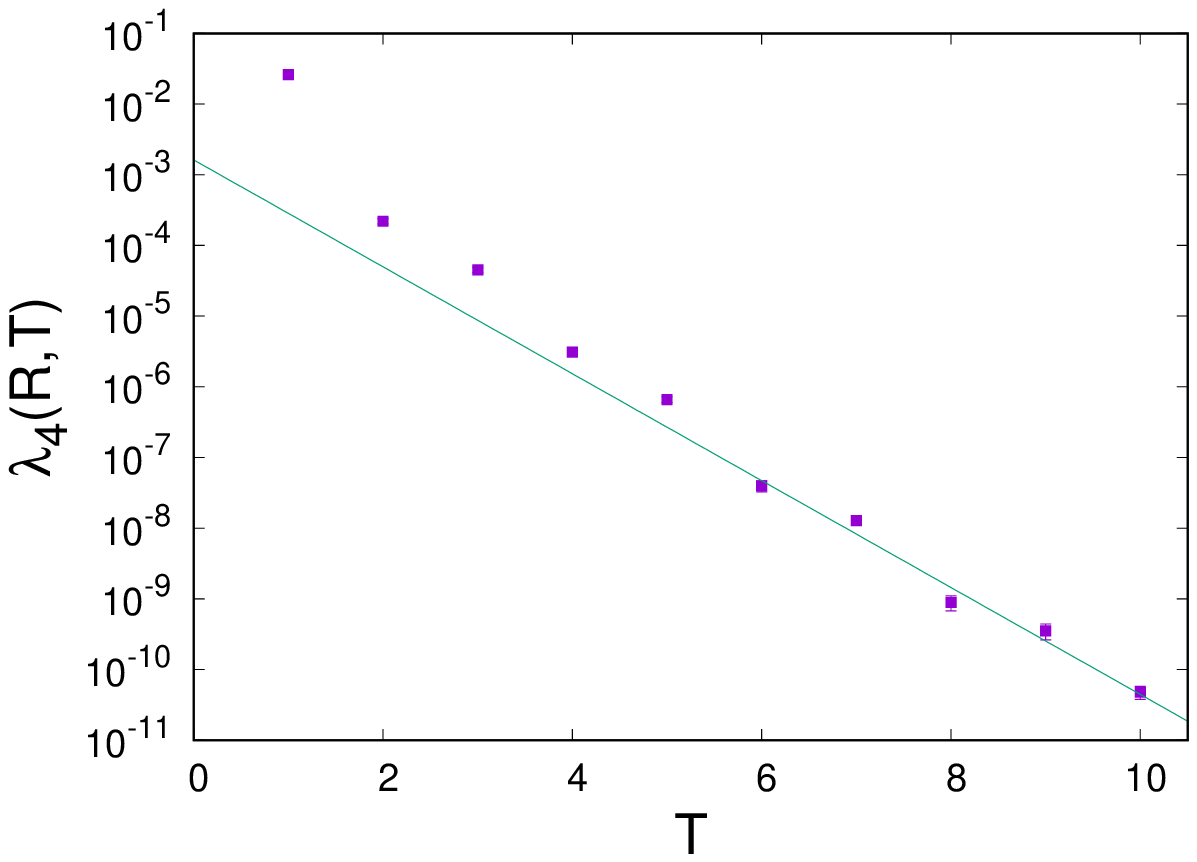}
\label{rx2g10}
}
\caption{Data points for $r=0$ (naive fermions) at $\g=1.0$. As in all previous plots, $R=4.243$. (a) $\l_1(R,T)$ vs.\ $T$.  
(b) $\l_2(R,T)$ vs.\ $T$.  The straight line is only to guide the eye. (c) $\l_2(R,T)$ vs.\ $T$ with a fit through the first four data points. (d) $\l_4(R,T)$ vs.\  $T$, with a fit to the last five data points.}
\label{rxg10}
\end{figure}

    It is also interesting to set $r=0$, just to compare with the previous results.  Of course with $r=0$ this U(1) theory is not chiral
in any sense because of the doublers.  There must now be left and right-handed fermion components which are degenerate in mass, and have the same abelian  charge.  As before, the ground state energy is easy to compute from $\l_1(R,T)$, as seen in
Figure \ref{rx1g10}.  However, there is an obstacle to obtaining accurate estimates for the first excited states, which is apparent in the data for $\l_2(R,T)$ shown in Fig.\ \ref{rx4g10}, where the straight line is only to guide the eye. Note that the data in this figure seems to oscillate around some average straight line.\footnote{Such oscillatory behavior in Euclidean time is known to be characteristic for operators associated with staggered fermions, and naive fermions are equivalent to a set of staggered fermions.  Perhaps this accounts for the oscillations seen here.  I thank Maarten Golterman for this remark.} We can extract an excited state from the
first four data points (Fig.\ \ref{rx3g10}), noting that the best fit extrapolates very close to 1 at $T=0$.  The oscillations are somewhat
less pronounced in $\l_4(R,T)$, as shown in Fig.\ \ref{rx2g10}, and we can attempt a best fit through the last five data points,
ignoring the very large $\chi^2$ values and error bars that result.  The energies obtained by this procedure are shown
in Fig.\ \ref{ERrxg10}, with the convention that the $E_n$ are ordered by increasing energy, whether or not $E_1$ corresponds
to $\l_1$.  What is striking is that the lowest energies seem to lie {\it below} the energies obtained from $\l_1$, suggesting that the lowest energy state is not simply the state $\Phi_9$ obtained by neutralizing charge with the Higgs field.
On the other hand, the error bars on these lower energies are huge, as is the scatter in the data, and we may question whether this result is even qualitatively trustworthy.  For this reason it is worth reexaming the situation at a different point in the phase diagram.
 
 \subsection{Higgs phase at $\b=3, \g=0.5$}  

   The point in the phase diagram at $\b=3,\g=1$ is very deep in the Higgs phase.  Although we do not attempt in
this paper to explore all regions of the Higgs phase, it is still of interest to carry out this investigation at a point which
is a little closer to the massless-to-Higgs transition line.  The transition to the massless phase at $\b=3$ occurs at $\g=0.32$,
and we will study a point a little above that transition point, at $\g=0.5$.  We recompute the photon mass in the pure
gauge Higgs system by the means described earlier, arriving at $m_{ph}=0.324(7)$ in lattice units.

 \subsubsection{Wilson term WM1}
 
\begin{figure}[htbp]
\includegraphics[scale=0.6]{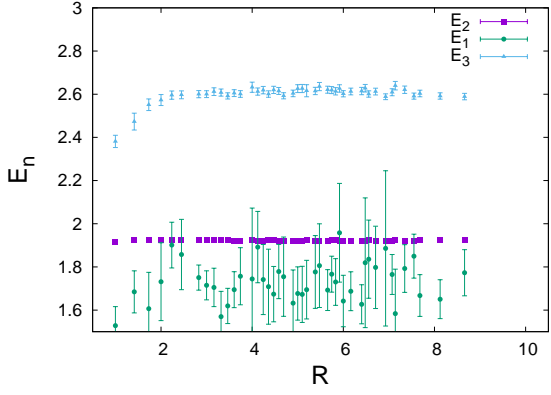} 
\caption{Energies for $r=0,\g=1.0$.}
\label{ERrxg10}
\end{figure}

\begin{figure}[htpb]
 \subfigure[~]{ 
 \includegraphics[scale=0.5]{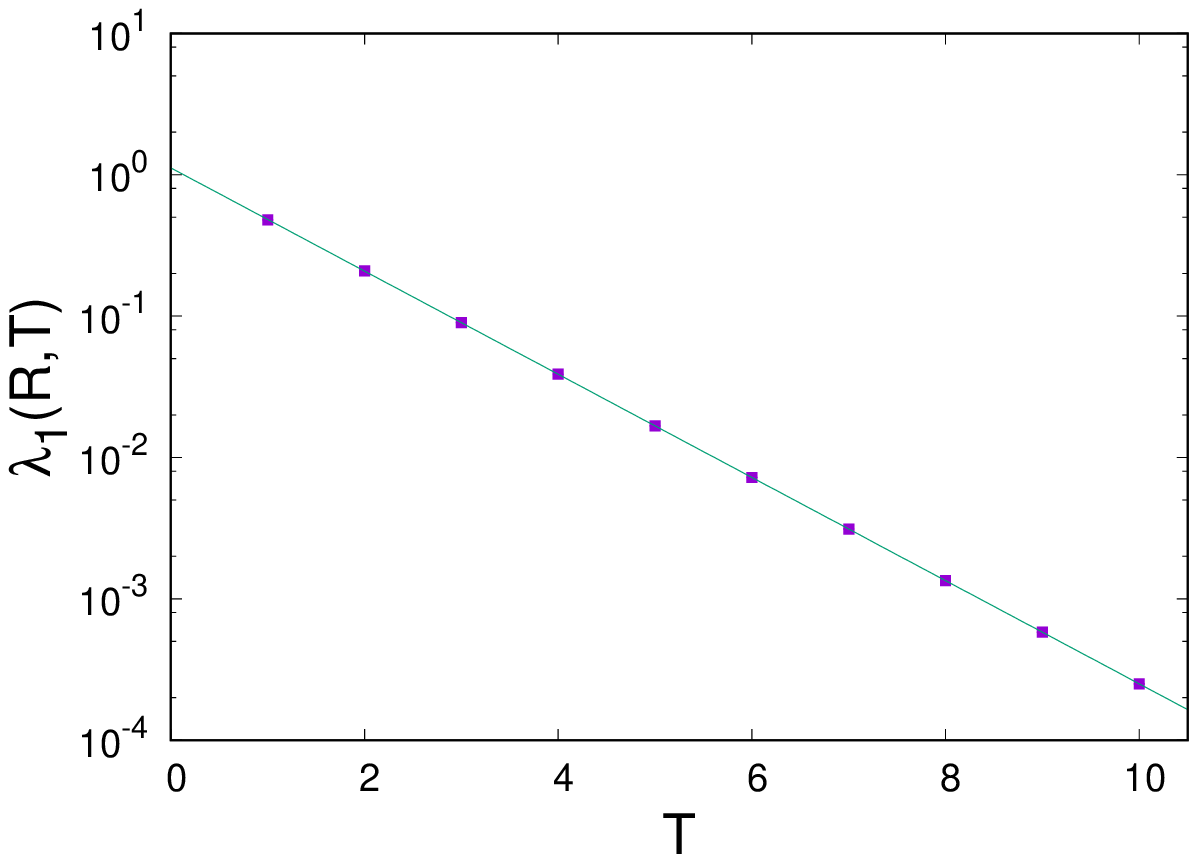}
\label{zx1g05}
}
 \subfigure[~]{  
 \includegraphics[scale=0.5]{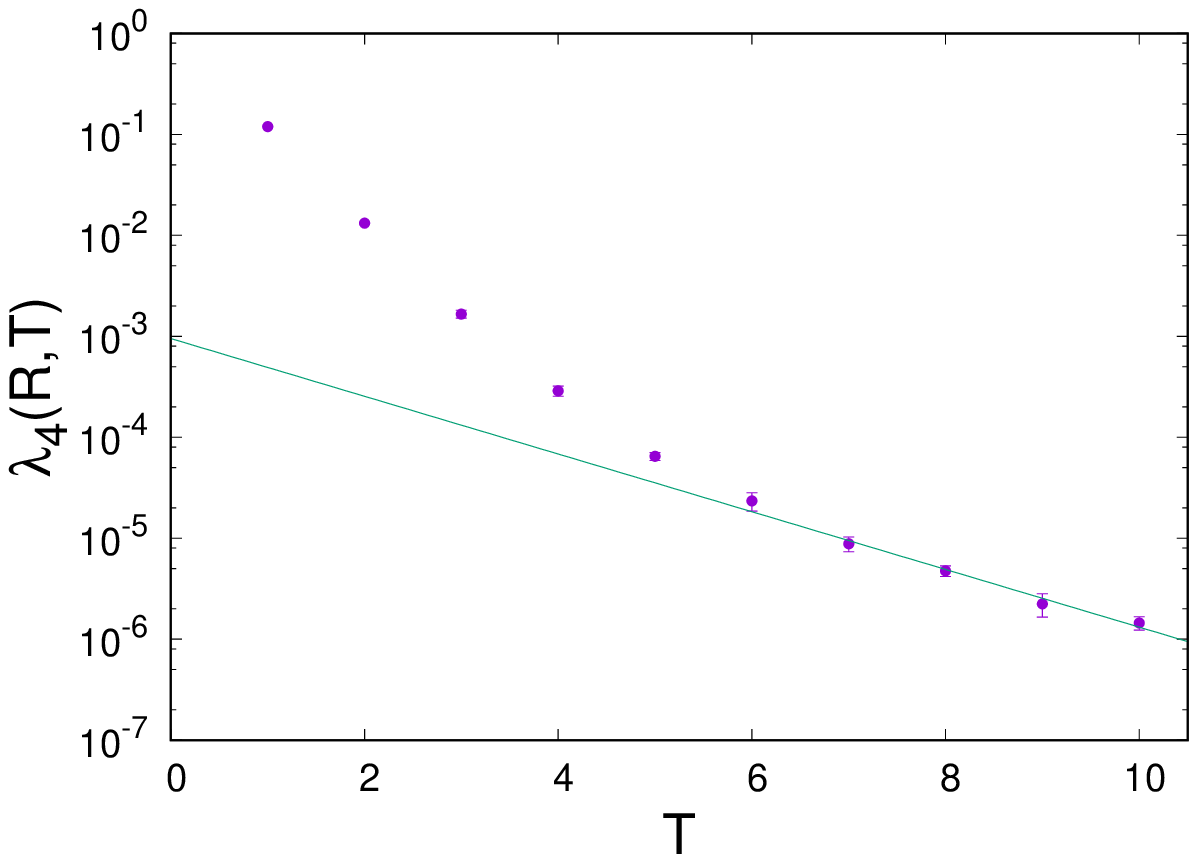}
\label{zx2g05}
}
 \subfigure[~]{  
 \includegraphics[scale=0.5]{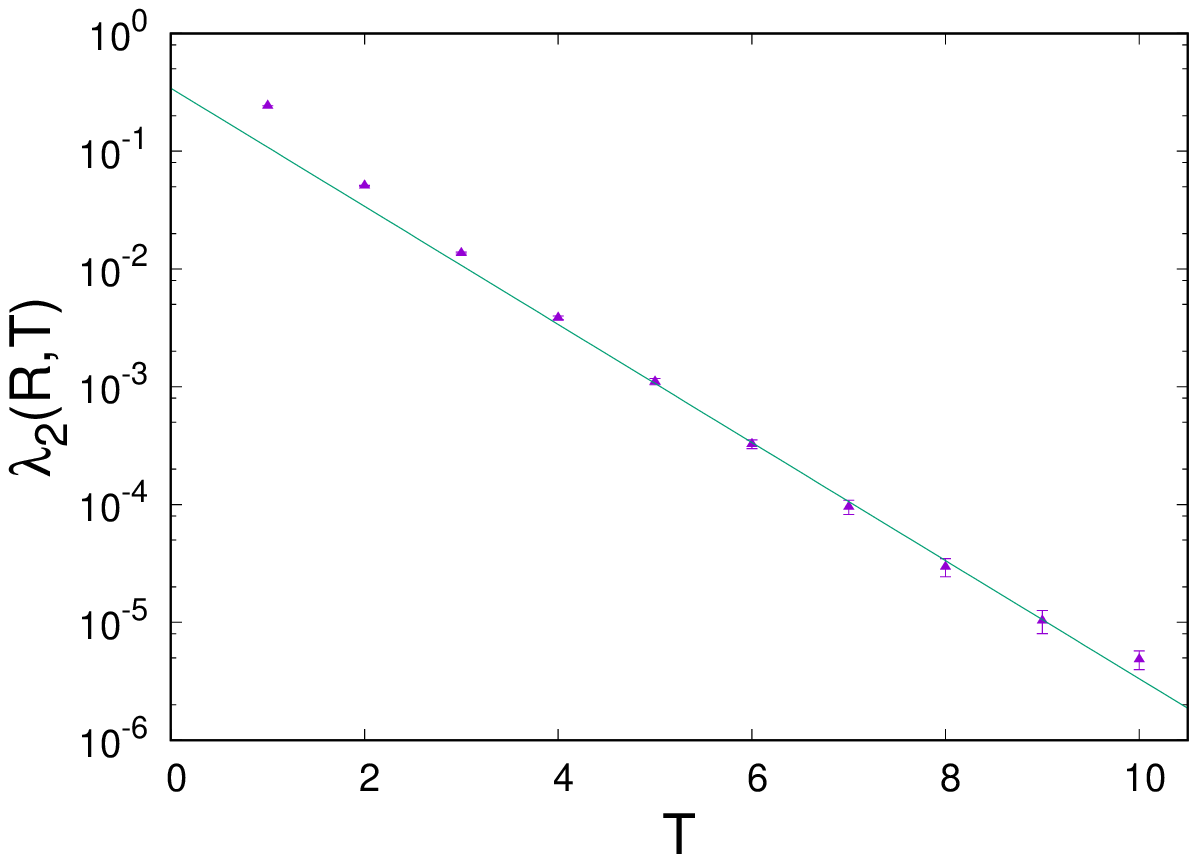}
\label{zx3g05}
}
\caption{ Data for the WM1 Wilson term at $\b=3.0, \g=0.5$. \newline  (a) $\l_1(R,T)$ vs.\ $T$.  Energy $E_2$ is extracted from this fit.  (b) $\l_4(R,T)$ vs.\ $T$.  Energy $E_1$ is extracted from this fit.  (c) A higher energy $E_3$ is extracted 
from a fit to the $\l_2(R,T)$ data in the range $T=5-10$.}
\label{zxg05}
\end{figure}

     Once again, the state $\Psi_1(R,T)$ seems to be very close to an exact eigenstate of the full transfer matrix, as is
evident from the plot of $\l_1(R,T)$ shown in Fig.\ \ref{zx1g05}.  In this case, however, $\Psi_1$ is not entirely the neutral
state $\Phi_9$.  Instead, the squared overlap \rf{f} is $f \approx 0.89$ (at all $R,T$), indicating a non-negligible mixture with pseudomatter fields.  Again we extract energies from plots of $\l_4(R,T)$ and $\l_2(R,T)$ (Figs.\ \ref{zx2g05} and \ref{zx3g05} ).  But here we again find evidence that the energy eigenvalues
 extracted from $\l_1(R,T)$ are actually not the ground state energies of the system at any $R$.  The energies derived from the large $T$ behavior of $\l_4(R,T)$ are consistently a little below those values.  There is a lot of scatter in our estimate of those energies,
 but it is still an improvement over what is seen Fig.\ \ref{ERrxg10}. Changing the fitting interval from $T=7-10$ to, e.g., $T=6-9$, affects those estimates slightly, but the energies remain below the points labeled $E_2$.  Despite the scatter in the data, it is
 clear that the energy separations between $E_1(R)$ and $E_2(R)$ are in general less than the photon mass, so the first
 excited state is again stable.
 
    One may ask how the energy derived from $\l_4$ can be less than that extracted from $\l_1$.  The answer is that
the index $n$ refers to the magnitude of the $\l_n$ in descending order, so that by definition $\l_1 > \l_4$, but the energies are derived from the slope of the $\l_n$ in some range on a log plot.  It can happen that in some range of $T$ the slope of $\l_4$ on a log plot is larger (i.e.\ less negative) than that of $\l_1$, despite the fact that $\l_1>\l_4$, and this is in fact clear from comparing Figs.\ \ref{zx1g05} with \ref{zx2g05}.  Of course it must ultimately happen, as $T$ increases, that data with the less negative slope will cross the data with the more negative slope, which would necessitate a relabeling of the $\l_n$ beyond the crossing point, but this is unnecessary in the range of $T$ that has been investigated.

    I have also looked at the WM2 case, and in this case the ground state still appears to be $\Psi_1$.  But the $\chi^2$ values, error bars, and scatter for the energy of the first excited state are very large, comparable to what is seen in Fig.\ \ref{ERrxg10}, and, like the data in that figure, must be regarded as untrustworthy.  

 \begin{figure}[htbp]
\includegraphics[scale=0.5]{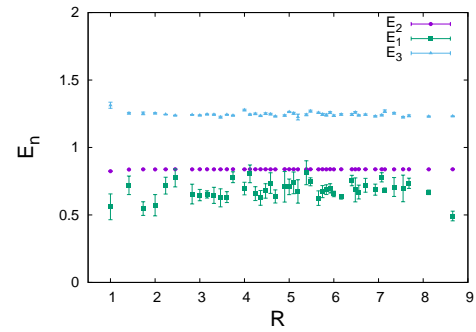} 
\caption{Energies $E_1,E_2,E_3$ vs.\ $R$ at $\b=3, \g=0.5$, shown together with the one photon threshold.  The Wilson mass term is WM1.  Note that the energies derived from $\l_4$ lie below the energies from $\l_1$.}
\label{ERzxg05}
\end{figure} 
 
 \begin{figure}[htpb]
 \subfigure[~]{
 \includegraphics[scale=0.5]{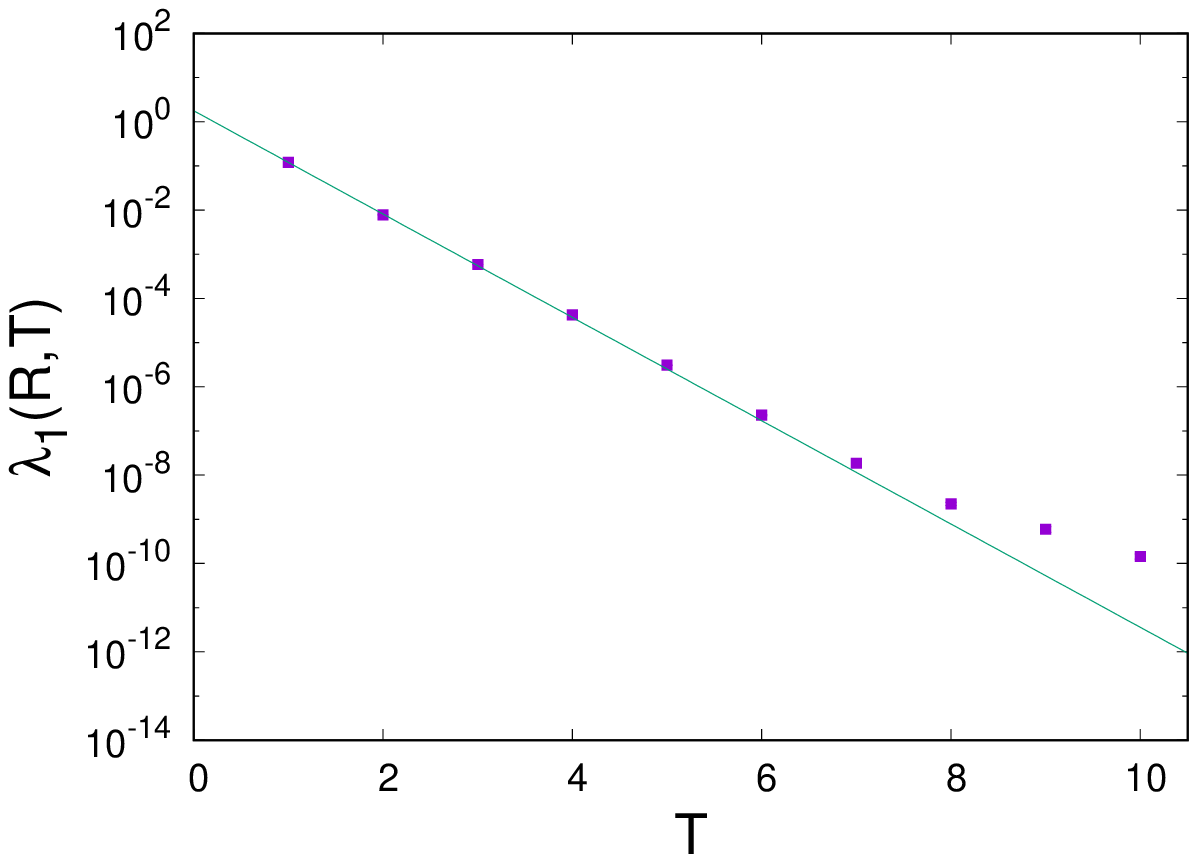}
\label{rx1g05}
}
\subfigure[~]{
 \includegraphics[scale=0.5]{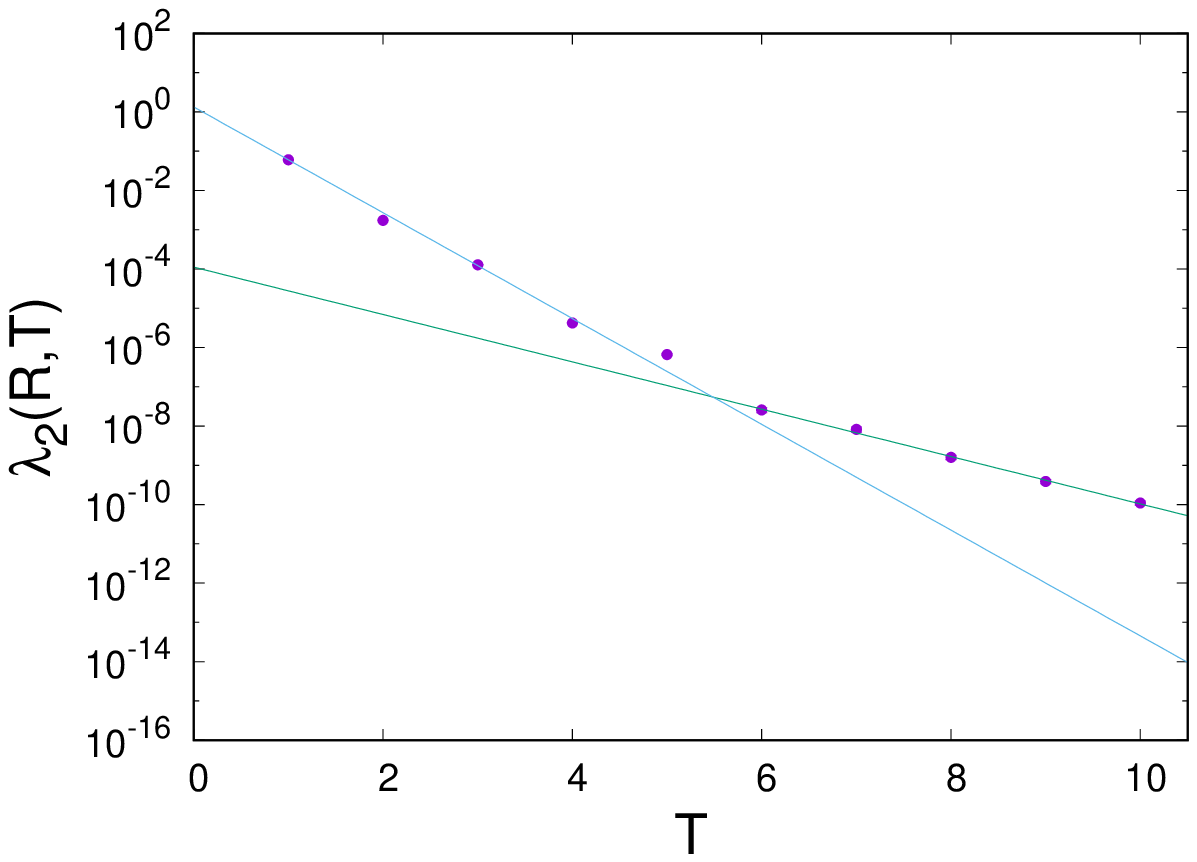}
\label{rx2g05}
}
\caption{$r=0$ data at $\b=3,\g=0.5.$  (a) $\l_1(R,T)$ vs.\ $T$.  Energy $E_2$ is extracted from this fit in the range $T=1-6$.  Notice that, unlike previous cases, the data points bend away from the straight line at large $T$.  (b) Energies $E_3$ and $E_1$ are both extracted from fits to the $\l_2(R,T)$ data in the ranges $T=1-4$ and $T=6-10$ respectively.}
\label{rxg05}
\end{figure} 
 
 \begin{figure}[htbp]
\includegraphics[scale=0.6]{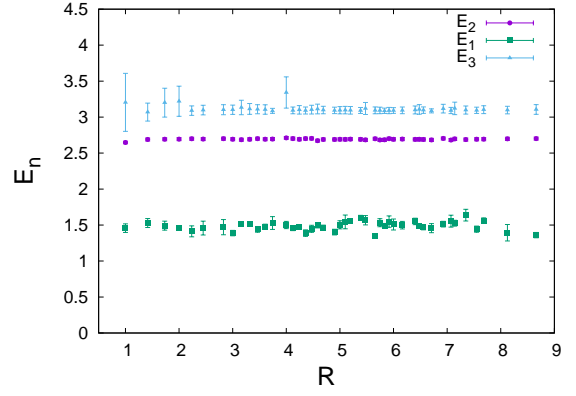} 
\caption{Energies $E_1,E_2,E_3$ vs.\ $R$ at $\b=3, \g=0.5$,  for the $r=0$ case with no Wilson mass term.  Here again,  the
energies extracted from fits to $\l_1$ are not the ground state energies.}
\label{ERrxg05}
\end {figure}

 \subsubsection{no Wilson term, $r=0$}
 
    The phenomenon seen in Figs.\  \ref{ERrxg10} and \ref{ERzxg05}, where the ground state energy seems to lie a little below the energy eigenvalue deduced from $\Psi_1$, is very evident in the $r=0$ case at $\g=0.5$.  This fact is already seen from
the plot of $\l_1(R,T)$ shown in Fig.\ \ref{rx1g05}, where the deviation of the last few data points from a straight line fit is an indication of the presence of a state of lower energy than the energy extracted from the slope of that line.  We can 
derive both a lower and a higher energy eigenvalue from fits to the large and small $T$ data, respectively, of $\l_2(R,T)$
(Fig.\ 10(b)).  The oscillation seen for $r=0$ at ${\g=1}$ is still there, although it is much less pronounced than in the data for
$\l_{n>2}$.  The three energy eigenvalues determined from the $\l_1, \l_2$ data are shown in Fig.\ \ref{ERrxg05}.  We note that the energy $E_1$ can also be determined from the large-$T$ data of $\l_3$ and $\l_4$, and these values are consistent
with the ground state energy determined from $\l_2$.  Since the energy eigenvalues $E_{2,3}$ exceed $E_1 + m_{ph}$, they
would be consistent with a combination of ground state + massive photon or, conceivably, a long-lived metastable state. \\

 \section{Conclusions}\label{rx2g05}
   U(1) chiral gauge Higgs theory does not seem to be of much physical interest in its own right. The intention in this article was to gain experience in applying the methods of refs.\ \cite{Greensite:2020lmh} and  \cite{Kazue} for vector-like theories, specifically SU(3) gauge Higgs and the $q=2$ abelian Higgs model respectively, to a simple gauge Higgs theory with static chiral fermions, and no fermion loops.  Even with non-dynamical fermions, and even with a simple U(1) gauge group, there are substantial complications as compared to the vector-like theories.  But in the end we find a similar result:  there is at least one stable excitation in the Higgs phase of a static fermion-antifermion pair, irrespective of pair separation.  Obviously the real question of interest is whether a result of this kind will be found in the electroweak sector of the Standard Model.  Perhaps a more sophisticated approach to lattice
chiral fermions in the electroweak sector, e.g.\ \cite{Kadoh:2007xb,*Kadoh:2008zz}, could be applied.  Such an approach, involving the lattice overlap operator, would be numerically challenging even in the quenched approximation (certainly by comparison to what we have done so far), but perhaps the calculation is feasible.  I hope to return to this question in a subsequent publication.
 
\bigskip

\acknowledgments{I would like to thank Maarten Golterman for helpful discussions. This research is supported by the U.S.\ Department of Energy under Grant No.\ DE-SC0013682.}

\bibliography{sym3}

\end{document}